# Nano-engineering of electron correlation in oxide superlattices

J. Laverock[1,2,*], M. Gu[3], V. Jovic[4], J. W. Lu[5], S. A. Wolf[3,5], R. M. Qiao[6], W. Yang[6] and K. E. Smith[1,4]

[1] *Department of Physics, Boston University, 590 Commonwealth Avenue, Boston, MA 02215, USA*
[2] *H. H. Wills Physics Laboratory, University of Bristol, Tyndall Avenue, Bristol BS8 1TL, UK*
[3] *Department of Physics, University of Virginia, Charlottesville, VA 22904, USA*
[4] *School of Chemical Sciences and MacDiarmid Institute for Advanced Materials and Nanotechnology, University of Auckland, Auckland 1142, New Zealand*
[5] *Department of Materials Science and Engineering, University of Virginia, Charlottesville, VA 22904, USA*
[6] *Advanced Light Source, Lawrence Berkeley National Laboratory, Berkeley, California, CA 94720, USA*

**Abstract**

Oxide heterostructures and superlattices have attracted a great deal of attention in recent years owing to the rich exotic properties encountered at their interfaces. We focus on the potential of tunable correlated oxides by investigating the spectral function of the prototypical correlated metal $SrVO_3$, using soft x-ray absorption spectroscopy (XAS) and resonant inelastic soft x-ray scattering (RIXS) to access both unoccupied and occupied electronic states, respectively. We demonstrate a remarkable level of tunability in the spectral function of $SrVO_3$ by varying its thickness within the $SrVO_3/SrTiO_3$ superlattice, showing that the effects of electron correlation can be tuned from dominating the energy spectrum in a strongly correlated Mott-Hubbard insulator, towards a correlated metal. We show that the effects of dimensionality on the correlated properties of $SrVO_3$ are augmented by interlayer coupling, yielding a highly flexible correlated oxide that may be readily married with other oxide systems.

**Introduction**

Recent advances in layer-by-layer epitaxial growth have brought oxide heterostructures, interfaces and superlattices into the spotlight in the search for promising emergent physics that is both technologically useful and eminently tunable.[1] The bringing together of two or more otherwise different materials with sometimes disparate properties can yield unexpected and exciting phases, the most well-known of which is the highly conducting two-dimensional electron gas found at the interface of two insulating materials, $LaAlO_3$ and $SrTiO_3$ (LAO/STO).[2] Of such oxide heterostructures, those involving *correlated* oxides are particularly appealing,[3] owing to the rich emergent phenomena that develops due to competing orbital, spin and charge degrees of freedom, e.g. superconductivity[4] or ferromagnetism[5] in LAO/STO. While a great number of studies have focused on the interface between band and/or correlated insulators[2,4-8] superconductors and/or ferromagnets[9] or with a view towards tailoring a material's orbital polarization,[10] there has been comparatively much less work on correlated metallic hetero-interfaces and superlattices.[11-13] In correlated light transition metal oxides, the 3$d$ band is dominated by the competition between the kinetic energy of the electrons (which scales with the bandwidth, $W$, and favours electron delocalization) and the on-site Coulomb repulsion, $U$ (which prefers to keep electrons apart and localized on atomic sites). The impact of strong electron correlations on materials properties can be profound, for instance leading to huge conductivity changes in the vanadium oxides. As electron correlations are increased, e.g. larger $U$ or smaller $W$, incoherent side bands develop either side of the one-electron-like quasiparticle peak (QP) in the spectral function, corresponding to localized electron states,[14] and referred to as lower and upper Hubbard bands (LHB and UHB). As a signature of electron correlations, both the energy and spectral weight of the Hubbard bands are important, as spectral weight is transferred from the QP into the Hubbard bands, which move progressively further apart in energy.[15] Eventually, the Coulomb repulsion is sufficient to fully localize the system, and it undergoes a Mott-Hubbard metal-insulator transition (MIT)[16] into a Mott insulator as the QP residue is fully depleted.

SrVO$_3$ (SVO) is a prototypical correlated metal,[17] exhibiting similar spectral weight in both its coherent QP and incoherent Hubbard sub-bands, see for example Figure 1d.[17-18] While bulk cubic SVO is a good correlated metal, ultrathin SVO has been observed to undergo an MIT as a function of thickness below ~ 2.7 nm.[19-20] As recently highlighted, this result is surprising,[21] and was originally attributed to the reduced dimensionality of the band structure leading to a reduction in the bandwidth of the active V $t_{2g}$ orbitals.[19-20] Detailed dynamical mean-field theory (DMFT) calculations refined this explanation as due to the lifting of the degeneracy of the $t_{2g}$ orbitals at the interface with STO, where the subsequent orbital redistribution triggers a Mott-Hubbard transition.[21] This emphatic electronic redistribution, triggered at the interface but occurring throughout the SVO layer, has been proposed as a high-performance Mott transistor,[21] providing a clear pathway towards exploiting SVO/STO in the fast-growing field of oxide electronics.[22]

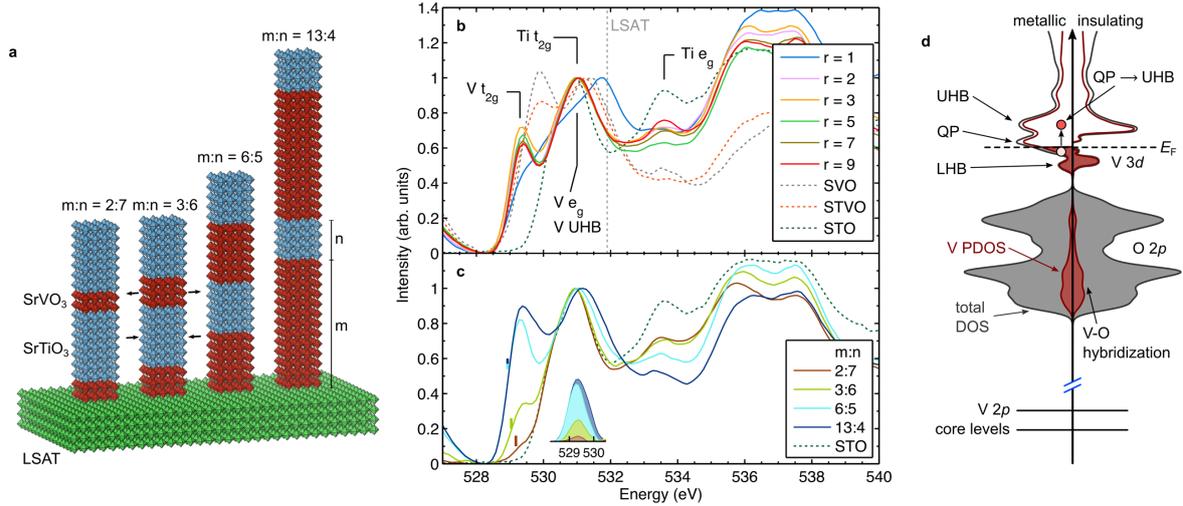

**Figure 1**. **Structure of [(SVO)$_m$/(STO)$_n$]$_r$ SLs and their electronic structure**. **a** Schematic illustration of the structure of the *mn*-series SLs. The arrows indicate the tensile and compressive in-plane strain experienced by the SVO and STO layers respectively. **b** O $K$ edge XAS of the *r*-series and **c** *mn*-series SLs, normalzed to the Ti $t_{2g}$ peak. Also shown are spectra of pure STO, pure SVO and solid solution SrTi$_{0.33}$V$_{0.67}$O$_3$ (STVO) for comparison. The location of the first LSAT peak is shown by the dotted line in **b**. The vertical bars in **c** indicate the leading edges of the spectra, and the inset shows a close-up view of the V $t_{2g}$ states after subtraction of the Ti $t_{2g}$ and V $e_g$ tails. **d** Schematic diagram of the correlated electronic structure of SVO, showing metallic and insulating densities of states.

Here, we report remarkable tunability in the degree of correlations in SVO/STO superlattices (SLs) as a function of the SL structure. The effects of correlation were measured using a variety of soft x-ray spectroscopies that allow us to probe the buried layers and interfaces (see Figure 1a). Specifically, we find that the SVO layers can be continuously tuned from a good correlated metal (in thick SVO layers) through a Mott-Hubbard transition into a strongly correlated Mott insulator (in thin SVO layers). We demonstrate that this is a collective effect which develops due to the coherence (through interlayer tunnelling) between the SVO layers in the SL. These results establish SVO/STO SLs as exemplar tunable correlated oxides, which may be epitaxially combined with other perovskite oxides in heterojunctions to build all-oxide electronic devices with designed properties.

**Methods**

The [(SVO)$_m$/(STO)$_n$]$_r$ superlattices were grown epitaxially on (001)-oriented (LaAlO$_3$)$_{0.3}$(Sr$_2$TaAlO$_6$)$_{0.7}$ (LSAT) substrates using a pulsed electron-beam deposition technique,[23-24] which employs a high-power pulsed electron beam source to ablate the target.

Additional (epitaxial) samples of pure SVO, pure STO and solid solution $Sr(Ti_xV_{1-x})VO_3$ were also grown in the same way (film thickness of 16 nm) on LSAT for comparison.[25] The use of a $Sr_2V_2O_7$ target in the PED growth, which has an excess O stoichiometry compared with $SrVO_3$, helps to minimize the formation of O vacancies during growth.[26] Single crystals of $Sr_{0.5}Ca_{0.5}VO_3$ and $YVO_3$ were measured as reference $V^{4+}$ and $V^{3+}$ perovskites. X-ray diffraction displayed strong Kiessig fringes for all films, indicating high quality superlattices with sharp, well-defined interfaces between SVO and STO.[27] The (average) out-of-plane lattice parameters (*c*) varied almost linearly with $m/(m+n)$, corresponding to Vergard's law, from 4.00 Å (*m*:*n* = 2:7) to 3.86 Å (*m*:*n* = 13:4). Atomic force microscopy of individual SVO and STO layers indicated atomically smooth surfaces with an r.m.s. surface roughness of less than 0.2 nm. Transport properties (electrical resistivity and magnetoresistance) were measured using the van der Pauw method.

XAS and RIXS measurements were performed at beamline 8.0.1 of the Advanced Light Source, Lawrence Berkeley National Laboratory. Throughout these measurements, samples were continuously monitored for changes during and following beam exposure and the x-ray flux was maintained at a level so as to avoid beam damage. XAS measurements, recorded in both total electron yield (TEY, sampling depth $\lambda_{TEY} = 4.1$ nm, see Supplementary Information) and total fluorescent yield (TFY, sampling depth $\lambda_{TFY} = 70.5$ nm, see Supplementary Information) were recorded with an energy resolution of 0.2 eV. Note that the depth sensitivity of XAS, particularly in TEY mode, is strongly dependent on the particular sample, since the top layers are each of different thicknesses (see Supplementary Information). At the V and Ti *L* edges, the TFY spectra were quantitatively similar to TEY (excepting saturation effects present in the TFY signal),[28] confirming the sensitivity of TEY is representative of the buried layers at these edges, rather than the film surface. In the manuscript, all XAS spectra presented were recorded using TEY, whereas TFY was used to characterize the depth sensitivity of the two measurements, and to ensure TEY was representative of the buried layers. Linear dichroism spectra were measured at x-ray incidence angles of 20° and 70° to the surface normal of the film, corresponding to 88% of the incident photon polarization being aligned along either the in- or out-of-plane directions. All XAS and RIXS spectra have been treated with Savitzky-Golay filtering to improve the signal-to-noise ratio,[29] except where explicitly stated. For clarity, we also present raw (unfiltered) RIXS spectra in the manuscript. The resolution of the emission spectrometer was set to 0.35 eV at full-width half maximum (FWHM). The sampling depth of the RIXS measurements is estimated to be similar to the TFY sampling depth, ~ 70.5 nm.

The electronic structure of the SLs was calculated using the QUANTUM ESPRESSO package,[30] using ultrasoft pseudopotentials within the generalized gradient approximation and cutoffs of 40 Ry and 400 Ry for the kinetic energy of the wavefunctions and charge density respectively. In-plane lattice parameters were clamped to the LSAT lattice parameter, $a_{LSAT}$ = 3.868 Å, and the out-of-plane parameters used were those determined by XRD: $c_{2:7}$ = 9 × 4.00 Å, $c_{3:6}$ = 9 × 3.97 Å, $c_{6:5}$ = 11 × 3.92 Å. The out-of-plane lattice parameter of bulk, strained SVO was that corresponding to the $m$:$n$ = 13:4 SL, $c_{strain}$ = $c_{13:4}$ / 17 = 3.86 Å.

**Results and Discussion**

Two series of [(SVO)$_m$/(STO)$_n$]$_r$ SLs are investigated here, see Figure 1a: one in which the number of repetitions, $r$, are varied for fixed $m$ = 7, $n$ = 4 (designated as an $r$-series), and one in which the ratio of SVO layers ($m$) to STO layers ($n$) is varied for $r$ = 9 (designated as an $mn$-series).[27] In-plane lattice parameters were clamped to the LSAT substrate lattice parameter ($a$ = $b$ = 3.868 Å), leading to in-plane tensile strain within the SVO layers ($a_{SVO}$ < $a_{LSAT}$), and in-plane compressive strain within the STO layers ($a_{STO}$ > $a_{LSAT}$), as illustrated in Figure 1a. For the $r$-series of SLs, metallic conductivity was observed for $r \geq 3$, whereas $r$ = 1 was found to be semiconducting. The $r$ = 2 SL exhibited a metal-insulator transition (MIT) at 115 K. Superlattices with $m \geq 7$ were metallic, whereas those with $m \leq 3$ were semiconducting; the $m$:$n$ = 6:5 SL had an MIT at 235 K.

X-ray absorption spectroscopy (XAS) at the oxygen $K$ edge (1$s$ → 2$p$) is closely related to the unoccupied partial density of states (PDOS) of transition metal oxides via 2$p$ – 3$d$ hybridization.[31] In Figure 1, O $K$ edge XAS spectra are shown of both $r$-series (Figure 1b) and $mn$-series (Figure 1c) SLs, alongside reference spectra of pure SVO, STO and solid solution SrTi$_{0.33}$V$_{0.67}$O$_3$ (STVO). As illustrated by the STO spectrum, unoccupied Ti $t_{2g}$ and $e_g$ states are found at 531 and 533.6 eV, above which are the higher energy Ti and V 4$sp$ states; there is no spectral weight below the Ti $t_{2g}$ states in pure STO. For the SLs, however, a strong peak develops at 529.3 eV due to the unoccupied V $t_{2g}$ band, which forms the metallic states near the Fermi level ($E_F$), and grows in intensity with the V content of the SL (Figure 1c). This peak is found at a slightly higher energy in pure SVO and solid solution STVO, along with a second peak at ≈ 531.4 eV, which is due to overlapping contributions from the V $e_g$ states and

V UHB.[32] The unoccupied electronic structure of the SLs is clearly quite different to either solid solution STVO or a weighted average of STO and SVO (even accounting for energy shifts due to the alignment of $E_F$ at the interface), underlining that these layered SLs have well-defined interfaces and layers (see Methods). For example, in Figure 1b, the spectrum of solid solution STVO with a Ti concentration of 0.33 is shown alongside the spectra of the *r*-series SLs, for which the average Ti concentration is 0.36.

The *r*-series XAS spectra are shown in Figure 1b, and are very similar for $r \geq 2$, though there exists a small, but discernible, progressive upshift (by ~ 0.1 eV) in the energy of the UHB peak with increasing *r*. However, for *r* = 1, i.e. a single two-dimensional (2D) layer of SVO just 2.7 nm thick capped with STO, the spectrum is quite different. In particular, the spectral weight near $E_F$ is weak and is shifted towards higher energies. Ultrathin SVO has been established to be insulating at thicknesses below 3 nm,[19-20] consistent with the results of Figure 1b. Alternatively, strongly correlated quantum well sub-bands have been observed for SVO thicknesses of between 1.9 and 3.8 nm,[33] which may explain the unusual PDOS of the *r* = 1 film, where the two broad slopes between 528.5 and 531.5 eV may reflect unoccupied quantized sub-bands of the quantum well. For $r \geq 2$, metallic behaviour is observed, both in the transport properties[27] and in our XAS spectra, despite no change to the layer structure, indicating that interlayer coupling (tunnelling) through the 1.6 nm thick STO layers is sufficient to overcome the effects of low-dimensionality that would otherwise lead to insulating behaviour.[34]

The unoccupied electronic structure exhibits a much greater sensitivity to the individual SVO and STO layer thicknesses, shown in Figure 1c for the *mn*-series. The spectrum of the STO-rich SL (*m*:*n* = 2:7) strongly resembles pure STO, with an additional small pre-edge peak due to V $t_{2g}$ electrons. Replacing just one of the STO layers with SVO (*m*:*n* = 3:6) leads to a modest increase in the sensitivity of the measurement to the SVO layers (from 16% to 24%, see Supplementary Information), but a huge increase in the spectral weight of the V $t_{2g}$ QP states, by a factor of > 6. As the SVO layer thickness is increased, the QP spectral weight rapidly grows, until it is comparable to that of the solid solution by *m*:*n* = 13:4, an increase of 24 times from the *m*:*n* = 2:7 SL, despite just a fourfold increase in the SVO layer sensitivity.

Further analysis of these results provides insight into the evolution of the electronic structure of SVO as the dimensionality of the system progresses from strongly 2D (*m*:*n* = 2:7) towards 3D (*m*:*n* = 13:4). The leading edge of the QP states, characterized here as the location

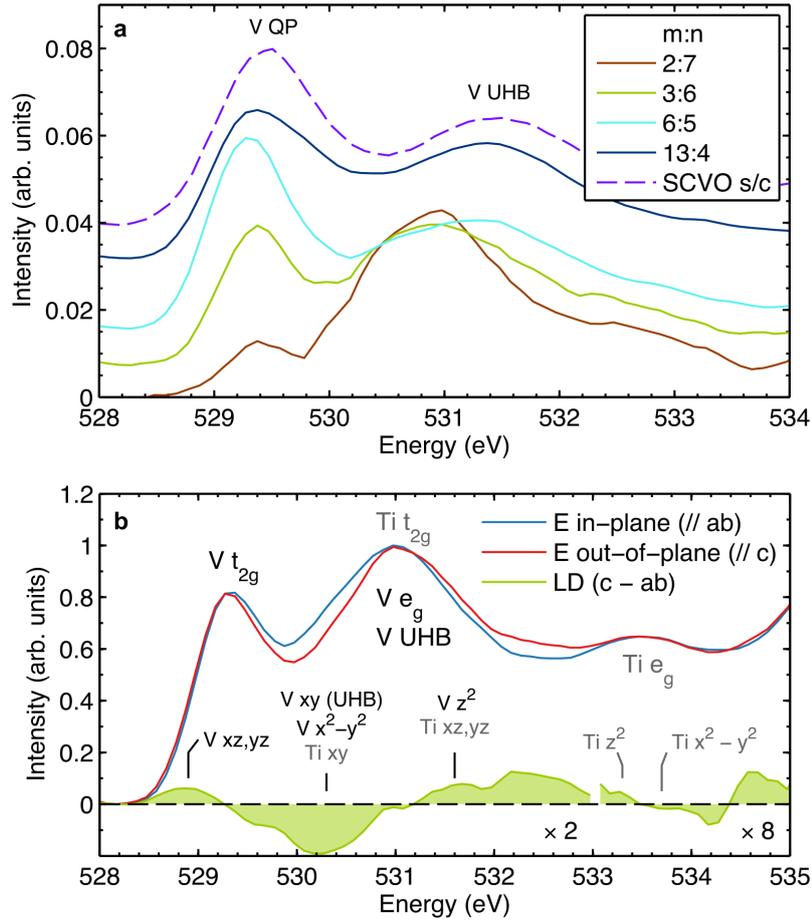

**Figure 2**. **Details of the SVO unoccupied electronic structure in the SLs**. **a** Extracted unoccupied V 3d PDOS of *mn*-series SLs, obtained by subtracting the STO contribution from the O K edge XAS spectra. The spectra have been normalized to the same area in the V 3d region. The spectrum of a $Sr_{0.5}Ca_{0.5}VO_3$ single crystal (SCVO) is also shown for reference. **b** Linear dichroism (LD) at the O K edge of $(SVO)_6/(STO)_5$. The two spectra were recorded with the polarization vector of the incident x-rays aligned with the SL plane ($\underline{E} \parallel ab$) and normal to it ($\underline{E} \parallel c$), to within 20°. The LD signal ($I_{\underline{E} \parallel c} - I_{\underline{E} \parallel ab}$) is shown amplified by a factor of 2 below 533 eV, and by a factor of 8 above in order to discern the weak LD in the Ti $e_g$ peak.

of the extrema in the first derivative, is shown as vertical bars in Figure 1c, and is 0.25 eV higher in energy for the 2D SL compared with the quasi-3D SL. Secondly, the bandwidth of the QP states, characterized as the full-width-half-maximum (FWHM) of the QP peak (see inset to Figure 1c), increases sharply with SVO layer thickness from 0.55 eV for *m:n* = 2:7 to 0.97 eV for *m:n* = 13:4. This increase in bandwidth (and the associated increase in metallicity) is accompanied by a transfer of spectral weight from the V UHB into the QP, typical of the evolution in the spectral function due to decreasing electron correlations.[14] In Figure 2a, the V 3d PDOS has been extracted from the *mn*-series spectra by subtracting the contribution of pure STO using the SVO/STO layer sensitivity weights given in the Supplementary

Information. These extracted spectra directly illustrate the evolution in bandwidth, spectral weight and energy of the UHB with SVO layer thickness. For example, for the 2D SL, just 9% of the spectral weight of the unoccupied V 3d states is located in the QP, whereas the QP and UHB are comparable in intensity for the single crystal and quasi-3D SL.

Linear dichroism (LD, $I_{\underline{E} \parallel c}$ - $I_{\underline{E} \parallel ab}$) can provide information on the energetic ordering of specific orbitals in low-symmetry systems.[5,35] In the bulk, SVO and STO are both cubic, and exhibit no LD. However, both the symmetry-breaking due to the interface and the epitaxial strain can introduce LD in our SLs. Representative LD of the SLs, reflecting the intrinsic orbital polarization of the electronic states, is shown in Figure 2b for $m:n$ = 6:5, for which TEY has a similar sensitivity to both SVO and STO layers (see Supplementary Information). The small positive LD at the leading edge indicates greater spectral weight from V $d_{xz,yz}$ orbitals lying just above $E_F$, suggesting the V $d_{xy}$ orbitals are preferentially occupied. This is in good agreement with DMFT predictions of few-layer SVO films,[21] but is difficult to separate from the effects of in-plane tensile strain, which can also lead to a lowering of the $d_{xy}$ orbital.[34] Although the Ti $t_{2g}$, V $e_g$ and V UHB overlap in energy, the rather strong negative dichroism below the 531 eV peak indicates that the bottom of the in-plane V $d_{x^2-y^2}$ and Ti $d_{xy}$ bands are lower in energy than the bottom of the out-of-plane V $d_{z^2}$ and Ti $d_{xz,yz}$ bands, similar to the effects of the CF distortion observed for TiO$_6$ octahedra in LAO/STO.[5,35] At higher energies, the V $d_{z^2}$ and Ti $d_{xz,yz}$ bands contribute a small positive LD. This interpretation and energetic ordering is supported by our DFT band calculations (see Supplementary Information).

Having measured the unoccupied electronic structure of the SVO/STO SLs, we now focus on the occupied states of the SVO layers through V $L$ edge resonant inelastic soft x-ray scattering (RIXS), which is capable of providing element-specific information on the local electronic structure,[36] and has been successfully employed in the measurement of buried layers and interfaces in a variety of systems.[37-40] Example RIXS spectra are shown in Figure 3a,b for the (SVO)$_7$/(STO)$_4$ SL. Fluorescent features are observed in Figure 3a due to the occupied PDOS of the system, including from hybridized V-O states and both incoherent (LHB) and coherent (QP) V $3d$ electrons. At intermediate excitation energies (e–i), RIXS features are well-separated from fluorescent emission, and $dd^*$-type excitations are visible at 2.1 and 3.3 eV, as illustrated in Figure 3b, consistent with previous measurements of single crystals.[32] We interpret the 2.1 eV transition as due to QP → UHB excitations (sketched in trapped at the interface, leading to quasi-localized charge distributed amongst neighbouring V sites; the electronic reconstruction at the interface will be discussed elsewhere.

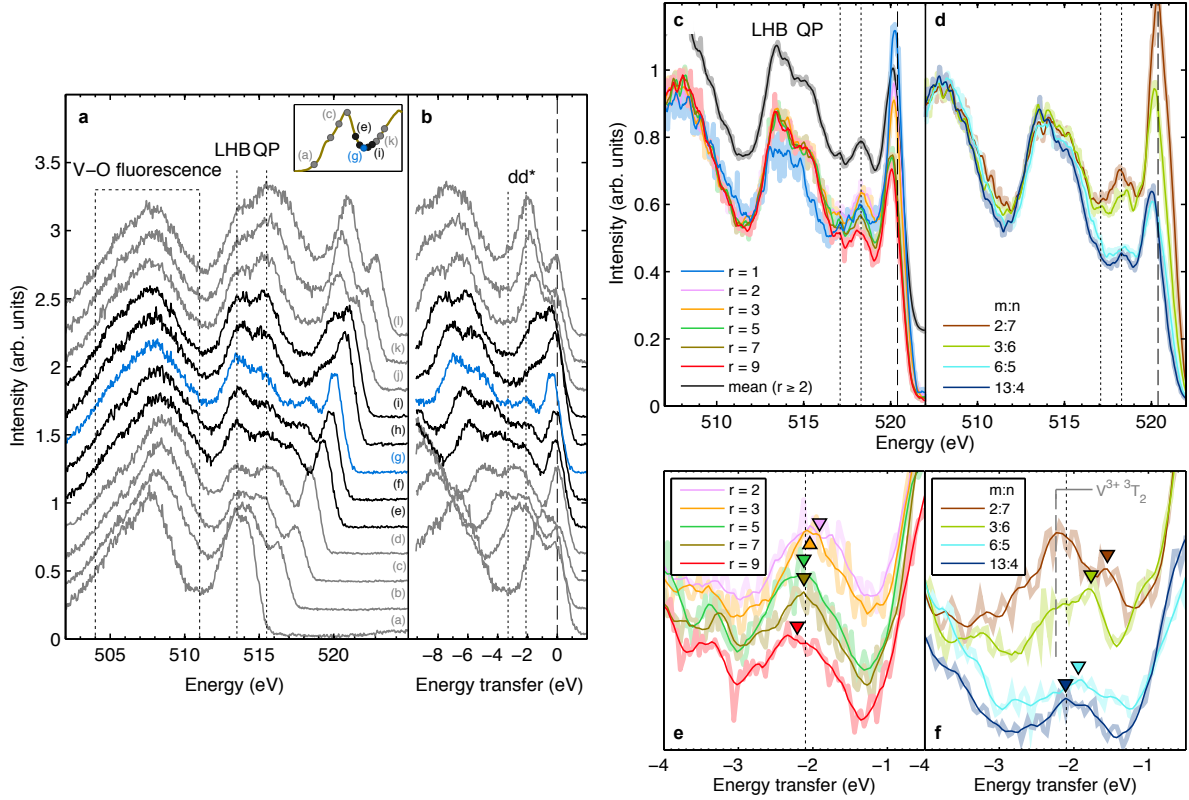

**Figure 3**. V $L$ edge RIXS of $[(SVO)_m/(STO)_n]_r$ SLs. **a** Example raw spectra for $m = 7$, $n = 4$ displayed on an emission energy scale, showing the location of the fluorescent features, and **b** an energy transfer scale, highlighting the RIXS features. Each spectrum (a–l) is recorded at a different RIXS excitation energy; the inset to **a** shows these excitation energies overlaid on the V $L$ edge XAS spectrum. Spectra (j–l) have been recorded near the $L_2$ edge and contain additional contribution from $L_2$ features. The blue spectrum is that chosen for further studies of the different SL structures. **c,d** RIXS spectra at excitation energy (g) of panel **a** for each SL: **c** $r$-series and **d** $mn$-series. The dashed lines in **c,d** indicate the elastic scattering energy, and the dotted lines illustrate the $dd^*$ transitions at 2.1 and 3.3 eV. For comparison, the mean of the spectra of the films (for $r \geq 2$) is also shown in **c**. **e,f** show a close-up view of the QP → UHB transition on an energy transfer scale for **e** the $r$-series SLs, and **f** the $mn$-series SLs. The vertical dotted lines in **e,f** represent the mean energy of 2.1 eV, and the triangles indicate the peak centers for each spectrum. The energy of $V^{3+}$ $^3T_2$ CF transitions is illustrated for the thin SVO layers in **f**. In panels **c-f**, the thick, lightly colored lines show the scatter of the unfiltered spectra.

Figure 1d), and that at 3.3 eV due to LHB → QP excitations. At low energy transfer ($\leq 1$ eV), metallic excitations (QP → QP) and intra-$t_{2g}$ transitions contribute broad energy scattering close to elastic scattering. As an exponent of all three types of feature in the RIXS spectra (V-O fluorescence, V $3d$ fluorescence and RIXS $dd^*$), spectrum (g) has been measured for each SL. We note that $V^{3+}$ species were detected at the interface in V $L$ edge XAS measurements (not shown here) at a concentration of ~ 0.1 per *interfacial* V ion. This weak charge doping of the interface may be associated with a small number of O vacancies that are energetically

The RIXS spectra of each SL are shown in Figure 3c,d. Although both *r*-series and *mn*-series spectra are rather similar in their fluorescence spectrum, fluorescent features in RIXS are generally rather broad, owing to the finite lifetime of the intermediate (core hole) state, making it difficult to untangle subtle modifications to the ground state. Consequently, we focus on the *loss* features of RIXS, namely the QP → UHB excitation, which are broadened only by the much narrower lifetime of the final state. The QP → UHB excitation is clearly visible at ~ 2.1 eV in all SL spectra. In Figure 3e,f we show the same spectra on an expanded view around the QP → UHB transition. For the *r*-series SLs ($r \geq 2$, Figure 3e), the transition energy increases by ~ 0.3 eV from 1.9 eV ($r = 2$) to 2.2 eV ($r = 9$). This evolution in the UHB energy was recognized in the XAS of the *r*-series above and was estimated at 0.1 eV, although the proximity of the Ti 3*d* states made it difficult to extract a reliable value. These results reinforce our proposition that increasing the number of repetitions, *r*, of SVO/STO layers decreases the impact of electron correlations by providing stronger interlayer coupling (via tunnelling) between SVO layers. For the *mn*-series, the presence of charge at the SVO/STO interface particularly affects thin SVO layers, where the ratio of interfacial V is high. In Figure 3f, an expanded view of the *mn*-series superlattices is shown, and a strong peak near 2.2 eV dominates the *m* = 2 spectrum, for which all V ions are located at the interface, and is also visible in the *m* = 3 spectrum. This peak can be associated with the additional interfacial charge in the form of $V^{3+}$-like RIXS, occurring at a similar energy to $^3T_2$ CF excitations found in $V^{3+}$ perovskites.[41] At lower energy transfer, a strong shoulder is visible in both of these spectra, and we associate this shoulder with the QP → UHB excitation discussed above. With this in mind, the evolution of this feature for the *mn*-series is remarkable, increasing by over 0.5 eV from ~ 1.5 eV for the 2D SL to ~ 2.1 eV for the quasi-3D SL. This is in very good agreement with the extracted XAS results, in which the energy of the UHB was also found to increase by 0.5 eV.

Taken together, our XAS and RIXS results reveal an intimate link between the thickness of the SVO layers in the SL and the correlated electronic structure. Moreover, this link is strongly influenced by the interlayer coupling, provided by tunnelling through the STO "spacer" layers. In order to demonstrate these effects in more detail, in Figure 4 we compile several important spectroscopic properties of the SLs from our XAS and RIXS results (see Supplementary Information). The first of these (Figure 4a) is the spectroscopic "metallicity" of the SLs, determined from the leading edge of the XAS (which represents the conduction band minimum for insulators and $E_F$ for metals) and the integrated intensity of quasi-elastic scattering in RIXS (which reflects the delocalization of the intermediate RIXS state).[42-43] As illustrated for the *mn*-series, both indicators of the spectroscopic metallicity of the SLs agree very well with their transport properties, whereas the *r*-series display only weak trends in both indicators. In particular, the absence of a large upshift in either quantity for the insulating $r = 1$ sample is in agreement with the presence of disconnected metallic and insulating phases implied by detailed transport measurements,[27] suggesting that the MIT that develops with $r$ is likely instigated by percolative transport through phase-separated domains. The RIXS indicator (which is of greater precision than the XAS indicator) implies that the SVO layers in the SL are fully metallic for r ≥ 5, whereas a fraction of insulating SVO may coexist with metallic SVO for r < 5.

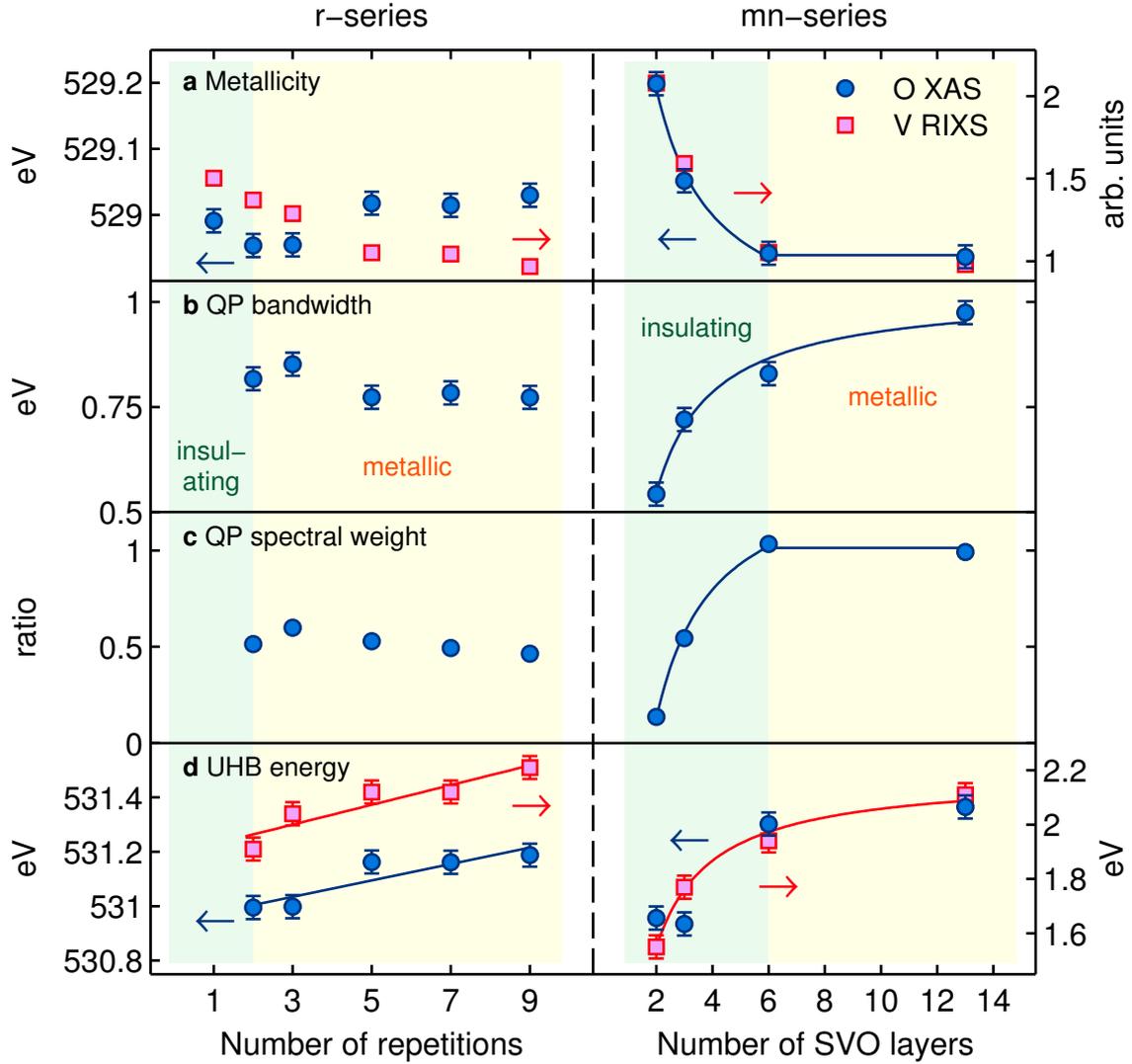

**Figure 4. Evolution in the electronic properties of SVO SLs** as a function of number of repetitions (*r*-series SLs) and number of SVO layers (*mn*-series SLs), determined by O *K* edge XAS and V *L* edge RIXS measurements. **a** Spectroscopic metallicity, determined by the energy of the leading edge (XAS) and the intensity of quasi-elastic scattering (RIXS). **b** QP bandwidth, determined as the FWHM of the QP in the XAS spectra. **c** Ratio of QP spectral weight to UHB spectral weight from XAS. **d** Energy of the UHB, determined as the maximum in the UHB peak in the extracted V unoccupied PDOS (XAS), and as the QP → UHB transition energy (RIXS). The guides to the eye represent linear (*r*-series) and $1/m$ (*mn*-series) fits to the data (see text). Error bars of ±0.05 eV (half the XAS point density) are shown for all energy measurements; all other error bars are smaller than the point size.

The remaining SL properties shown in Figure 4b-d are associated with the correlated electronic structure. The QP bandwidth is almost constant with $r \geq 2$, but exhibits a very strong dependence on the number of SVO layers, increasing by almost a factor of 2 from $m = 2$ to 13 (Figure 4b), and is well described by $1/m$ (best fit line in the right panel of Figure 4b). Figure 5 compares the electronic structure of the SLs calculated by density functional theory (DFT). Although DFT neglects the effects of strong electron correlation, the unrenormalized bandwidth, $W$, is directly accessible from DFT, and it is this quantity that competes with the on-site Coulomb energy, $U$ (which should not change with SL structure), in strongly correlated materials. As shown in Figure 5e, the bandwidth of the out-of-plane $d_{xz,yz}$ orbitals is found to vary (also with $1/m$) by a factor of ~ 1.4 between $m = 2$ and the bulk, in good qualitative agreement with the spectroscopic results, whereas the in-plane $d_{xy}$ orbitals are relatively only weakly affected. In contrast to the percolative MIT discussed above, Figure 4b suggests a huge renormalization in the $U/W$ ratio, which characterizes the relative energy scales of localized ($U$) versus delocalized ($W$) electron behaviour. In response, the transfer of spectral weight from the QP (large $m$) to the UHB (small $m$) is huge, by an order of magnitude for the insulating SLs (Figure 4c). Concurrently, the UHB rapidly lowers in energy as the thickness is reduced, and both RIXS and XAS quantifications are also described well by $1/m$ (Figure 4d). These two signatures of increased electron correlations, coupled with the associated reduction in bandwidth that leads to them, establish the MIT that develops in the $mn$-series of SLs is of the Mott-Hubbard type. Similar, albeit weaker, effects have been observed at the surface of $Sr_xCa_{1-x}VO_3$ single crystals due to the narrowing of the bandwidth as a result of the decreased coordination at the surface.[44] In the SLs, the narrowing of the bandwidth is caused by the decreased dimensionality of the system, whereby the out-of-plane $d_{xz}$ and $d_{yz}$ orbitals are confined within the SVO layer (see Figure 5), and the localization of these states due to an increase in $U/W$ triggers the localization of the in-plane $d_{xy}$ states, as illustrated for a single SVO layer by Zhong et al.[21] The (weaker) evolution in the UHB energy with the number of repetitions illustrates that interlayer coupling between SVO layers is able to restore some out-of-plane coherence via tunnelling through the STO layers, providing an additional tuning parameter of the correlated electronic structure.

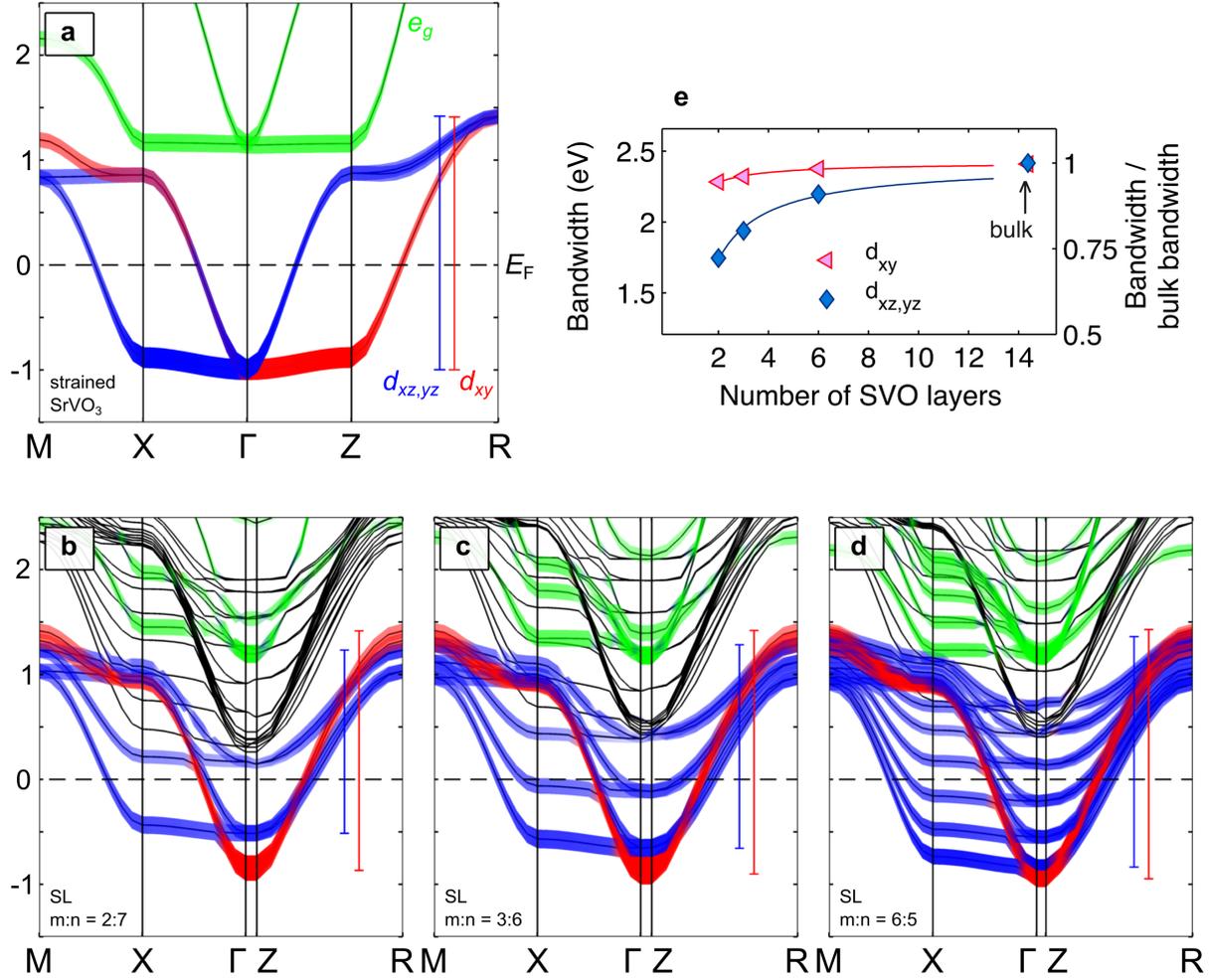

**Figure 5. Evolution in the SL bandstructure and bandwidths with SL structure.** Bandstructures of **a** strained, bulk SVO corresponding to the SVO lattice parameters of the $m:n = 13:4$ SL; **b-d** SVO/STO SLs: **b** $m:n = 2:7$, **c** $m:n = 3:6$, and **d** $m:n = 6:5$. In **a-d**, the V $d_{xy}$, $d_{xz,yz}$ and $e_g$ characters are shaded red, blue and green, respectively, and the corresponding bandwidths of the $d_{xy}$ and $d_{xz,yz}$ orbitals are indicated by the vertical bars. Unshaded bands correspond to Ti $3d$ bands. **e** The evolution in bandwidth of the $d_{xy}$ and $d_{xz,yz}$ states as a function of SVO layer thickness of the SLs, with the bulk (strained) values shown at the right. The solid lines represent $1/m$ fits to the data. The right axis indicates the normalized bandwidth for direct comparison with Figure 4b.

## Conclusion

Our results illustrate the remarkable flexibility in the correlated electronic structure of SrVO$_3$, and demonstrate that the superlattice structure is a sensitive tuning parameter that can be used to engineer both the material properties and electron correlations in SrVO$_3$, for example its proximity to the large Mott-Hubbard conductivity transition. This establishes superlattice engineering as a very attractive alternative to other approaches of tuning the electronic structure of correlated oxides. For example, strain engineering is limited in the range of available substrate mismatches, and crystallographic dimensionality approaches (e.g. the Ruddlesden-Popper phases) offer only coarsely-spaced distinct structures. On the other hand, superlattice engineering can be employed continuously from the 2D limit (ultrathin layers with a single layer or thick spacers) to the 3D limit (thick strongly-coupled layers). Moreover, as a perovskite oxide, SrVO$_3$ is widely compatible with other oxides, and promises an exciting role in the future study of emergent behaviour at correlated oxide interfaces.

**Acknowledgements**

The Boston University program is supported in part by the Department of Energy under Grant No. DE-FG02-98ER45680. The Advanced Light Source, Berkeley, is supported by the US Department of Energy under Contract No. DEAC02-05CH11231. M.G., J.W.L. and S.A.W. gratefully acknowledge financial support from the Army Research Office through MURI grant No. W911-NF-09-1-0398. Calculations were performed using the computational facilities of the Advanced Computing Research Centre, University of Bristol (http://www.bris.ac.uk/acrc/). We thank Stephen Dugdale for useful discussions.

# Supplementary information for "Nano-engineering of electron correlation in oxide superlattices"

J. Laverock[1,2,*], M. Gu[3], V. Jovic[4], J. W. Lu[5], S. A. Wolf[3,5], R. M. Qiao[6], W. Yang[6] and K. E. Smith[1,4]

[1] Department of Physics, Boston University, 590 Commonwealth Avenue, Boston, MA 02215, USA
[2] H. H. Wills Physics Laboratory, University of Bristol, Tyndall Avenue, Bristol BS8 1TL, UK
[3] Department of Physics, University of Virginia, Charlottesville, VA 22904, USA
[4] School of Chemical Sciences and MacDiarmid Institute for Advanced Materials and Nanotechnology, University of Auckland, Auckland 1142, New Zealand
[5] Department of Materials Science and Engineering, University of Virginia, Charlottesville, VA 22904, USA
[6] Advanced Light Source, Lawrence Berkeley National Laboratory, Berkeley, California, CA 94720, USA

## SA. Depth sensitivity of XAS in $SrVO_3/SrTiO_3$

The depth sensitivities of the two x-ray absorption spectroscopy (XAS) detection modes employed in this work (total electron yield, TEY, and total fluorescent yield, TFY) have been estimated from the relative contribution of the substrate (LSAT) signal to the recorded spectra for the various films of different thickness. This provides an internal measurement of the effective sample depths of TEY, $\lambda_{TEY}$, and TFY, $\lambda_{TFY}$. For TEY, which is much more surface sensitive, only the thinnest two samples (4.3 and 8.6 nm) available contained a detectable contribution from the substrate, LSAT, and we obtain $\lambda_{TEY}$ = 4.1 nm. For TFY, all samples (4.3 to 59.7 nm thick) exhibited a contribution from the LSAT substrate, which has been quantified as the ratio, $R$, of LSAT signal to SVO/STO superlattice (SL) signal in TFY.

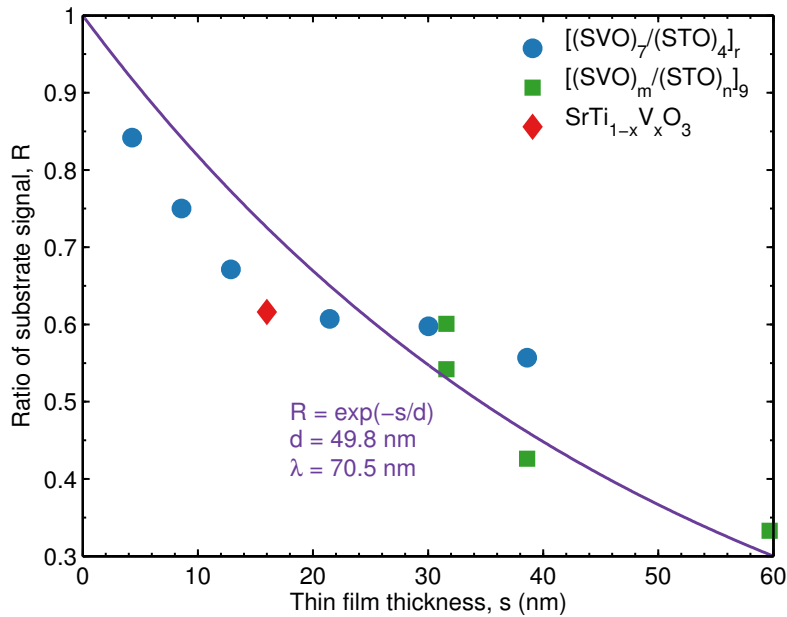

**Figure S1**: Relative contribution of substrate in TFY signal, plotted against film thickness. The exponent of the fit represents the path length of fluorescent x-rays.

This quantification is shown in Fig. S1, alongside an exponential fit to its dependence on film thickness. From this, we estimate that the intrinsic sampling depth of TFY is $\lambda_{TFY} = 70.5$ nm. In our measurements, the experimental geometry was set up so that the incident x-rays were $\theta_{inc} = 45°$ from the surface normal of the film; the depth sensitivity of our measurements in this geometry is $\lambda_{TFY} \cdot \cos \theta_{inc} = 49.9$ nm. Our measured values of $\lambda_{TEY}$ and $\lambda_{TFY}$ compare well with more accurate sampling depths obtained on other transition metal perovskites.[45]

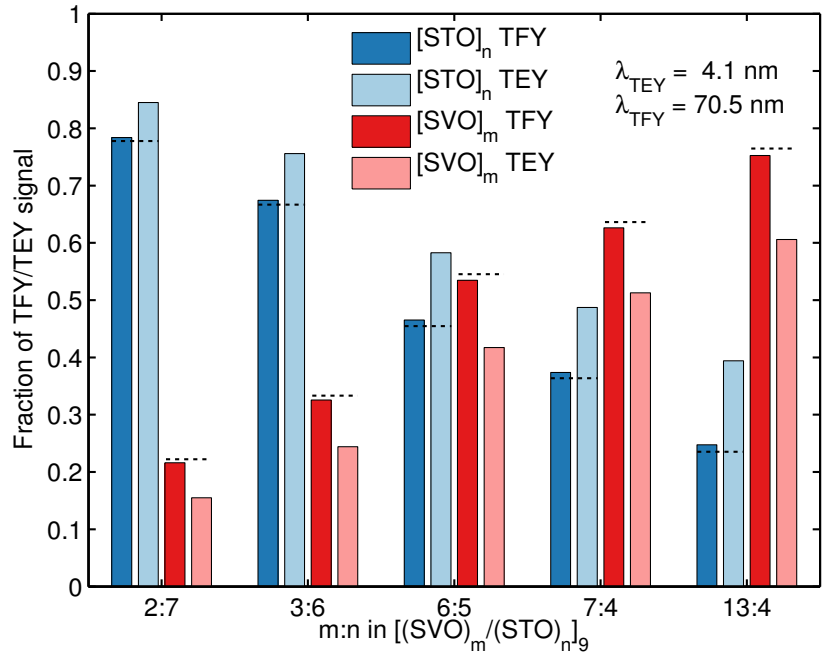

**Figure S2**: Total contribution of the SVO and STO layers in each SL to the TEY and TFY signals at the O $K$ edge. The dotted lines indicate the compositional fraction of the layers to the SL.

**SB. Sensitivity of TEY and TFY to individual SVO/STO layers at the O $K$ edge**

Owing to the different depth sensitivities of the two detection modes, the TEY and TFY signals contain different contributions from the SVO and STO layers for each SL. For clarity, this is illustrated graphically in Fig. S2, in which the relative contribution to both TEY and TFY of the SVO and STO layers are shown for each film (after subtraction of the LSAT contribution). As can be seen, the contribution to TFY more closely resembles the actual ratio $m / (m + n)$ for each film. On the other hand, TEY is more sensitive to the STO layers at the expense of the SVO layers due to the layer ordering of the SL (STO being the top layer) and its increased surface sensitivity.

**SC. Extraction of SL properties from XAS and RIXS data**

The SL properties presented in Fig. 4 of the manuscript were empirically determined from the O $K$ edge XAS and V $L$ edge RIXS measurements, using as few parameters as possible in the analysis. Below, we explain in more detail the rationale and method for each property.

(i) *Metallicity (O XAS)*. The most direct indicator of the metallicity of a system is the location of the first unoccupied states with respect to the Fermi level. The energies of the first unoccupied states of the SLs were determined from the leading edges of the raw O $K$ edge XAS data. The initial maximum in the first derivative of the raw O XAS was fitted locally to a Gaussian function, whose centre is interpreted as the leading edge. This method is parameter free. The best fit line to the *mn*-series data of Fig. 4(a) of the manuscript represents a $1/m$ function in the insulating region (between $m = 2$ and 6), and a constant value above $m \geq 6$ in the metallic region.

(ii) *Metallicity (V RIXS)*. The intensity of elastic scattering in RIXS is well known to be related to the delocalization of the intermediate state in the scattering process, and is therefore sensitive to the delocalized electrons, i.e. metallicity. Owing to the presence of low-energy metallic excitations (QP $\rightarrow$ QP transition) in the RIXS spectra of the metallic samples which are inseparable from elastic scattering within our resolution, our quantification also includes these kinds of excitations. All spectra were first normalized to the V-O fluorescence peak at ~ 508 eV in Fig. 3(a) of the manuscript. The intensity of elastic scattering was then determined by integrating each spectrum between 519.2 and 522.1 eV, i.e. at the minimum between quasi-elastic scattering and $dd^*$ transitions, and just above the elastic peak. The only parameters involved in this analysis is the choice of integration energies and normalization regions, and the results are robust against these choices.

(iii) *QP bandwidth (O XAS)*. The QP bandwidth was quantified from the O $K$ edge XAS data as the full-width at half-maximum (FWHM) of the QP, after subtraction of the STO spectrum. Although this does not represent the absolute bandwidth of the unoccupied QP states, as long as the shape of the QP PDOS is similar across the series (which is a reasonable assumption), it is proportional to it. For best results, one should obtain an accurate functional form for the $t_{2g}$ (QP), $e_g$ and UHB PDOS. However, given the complexity of the data, and the additional contribution from the STO layers at this edge, this is impractical. Instead, we subtract STO and V UHB tails from each spectrum such that the result remains positive in the region of interest [528 – 531 eV in Fig. 1(b)], approximated by the pure STO spectrum. We note that although this is a crude approximation, it should be reasonably reliable well away from the Ti $t_{2g}$ and V UHB states, and we consider it acceptable for our purposes. The FWHM is directly measured from the remaining spectrum, without any assumptions about the shape of the QP. In that sense, this approach is also parameter-free, but does contain some error in the approximation of the STO tails. The best fit line to the $mn$-series data of Fig. 4(b) of the manuscript represents a $1/m$ function.

(iv) *QP spectral weight (O XAS)*. The ratio of QP spectral weight to the UHB spectral weight has been determined from the extracted unoccupied V PDOS [Fig. 2(a)], based on the O $K$ edge XAS. The contribution from the STO layers (using the pure STO spectrum, and assuming the unoccupied electronic structure of STO does not evolve with the SL structure) was first subtracted from the raw spectra based on the weights given in Section SB to obtain an estimate of the unoccupied V PDOS of the SVO layers only. This is the same approach used to obtain the spectra in Fig. 2(a). The area under the QP was obtained by integrating the result between 528.4 eV and the energy of the minimum between the QP and UHB, which is different for each SL. The area under the UHB was obtained by integrating from this energy to 533.6 eV. The best fit line to the $mn$-series data of Fig. 4(c) of the manuscript represents a $1/m$ function in the insulating region (between $m = 2$ and 6), and a constant value above $m \geq 6$ in the metallic region.

(v) *UHB energy (O XAS)*. The UHB energy is determined from the same extracted V PDOS described above. The UHB energy is defined simply as the energy location of the maximum of the UHB feature, interpolated using quadratic interpolation between the peak channel and its neighbours either side. This method is parameter free.

(vi) *UHB energy (V RIXS)*. The UHB energy from V $L$ edge RIXS is defined as the QP → UHB transition energy. Owing to the presence of noise, and additional contributions from $V^{3+}$ $^3T_2$ excitations, this energy is located by eye, and is shown directly in Fig. 3(e,f). The best fit line to the $mn$-series data of Fig. 4(d) of the manuscript represents a $1/m$ function.

**SD. Ti character of bands and ordering of Ti orbitals**

Figure 5 of the manuscript shows the electronic band structure of the SLs labelled by the V $d$ partial character to illustrate the energetic ordering and evolution in the bandwidth of the V states in the SLs. To complement this, here we show the corresponding Ti $d$ partial characters in Fig. S3. Going from the $m:n = 7:2$ SL [Fig. S3(a)] to the 6:5 SL, the number of confined states decreases from 7 to 5. In all SLs, the $d_{xy}$ band has a larger bandwidth compared with the $d_{xz,yz}$ bands, similar to the V $d$ bands. Moreover, the band minima of the $d_{xy}$ bands are *lower* than those of the $d_{xz,yz}$ bands, similar to V and despite the compressive strain experienced by the STO layers.

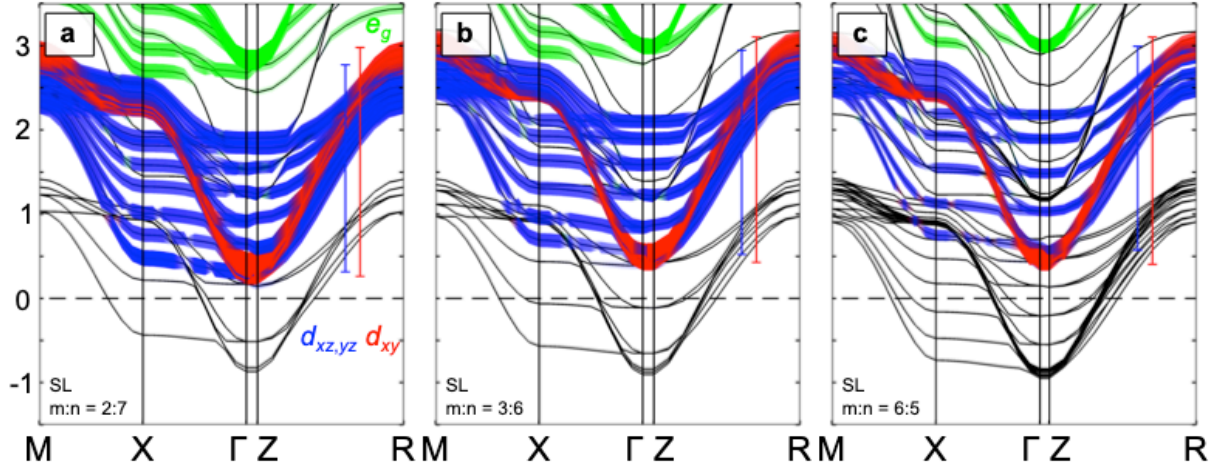

**Figure S3**: Electronic band structure of the SLs from DFT calculations (see manuscript for details). Bands are shown in the same way as Fig. 5 of the manuscript, except here the band character is labelled for the Ti $d$ character. In (a-c), the Ti $d_{xy}$, $d_{xz,yz}$ and $e_g$ characters are shaded red, blue and green, respectively, and the corresponding bandwidths of the Ti $d_{xy}$ and $d_{xz,yz}$ orbitals are indicated by the vertical bars. Unshaded bands correspond to V $3d$ bands.

Therefore, the lowest energy Ti $d$ states are $d_{xy}$ states, which are responsible for the small linear dichroism we observe in Fig. 3(b) of the manuscript, and which support our assignment of the energetic ordering of the Ti $t_{2g}$ states in the main manuscript. This is an electronic effect of the confinement in the $z$ axis of the SL, and is found in both the V and Ti states of the SVO and STO layers, respectively.

Within the Ti $e_g$ states, the linear dichroism we observe in Fig. 3(b) is reversed, implying that out-of-plane orbitals are lower in energy than in-plane orbitals, and that the $d_{z^2}$ orbital is the lowest available Ti $e_g$ state, as labelled in Fig. 3(b). Indeed, this observation, likely due to the compressive strain in the STO layers, is also supported by DFT; in Fig. S3, the $d_{z^2}$ bands (which are the lowest energy $e_g$ band at $X$) have a lower band minimum than the $d_{x^2-y^2}$ bands (whose minimum is at $\Gamma$) for each of the SLs.